\documentclass[twocolumn,showpacs,amsfonts,aps,prc,floatfix,superscriptaddress,longbibliography
]{revtex4-2}

\usepackage[colorlinks=true, pdfstartview=FitV, linkcolor=red, citecolor=blue, urlcolor=blue]{hyperref}
\usepackage{amsmath}
\usepackage{bm}
\usepackage{graphicx}
\usepackage{multirow}

\begin{document}

\title{
Signatures of QCD conductivities in heavy-ion collisions 
}

\author{Akihiko Monnai}
\affiliation{Department of General Education, Faculty of Engineering, \\
Osaka Institute of Technology, Osaka 535-8585, Japan}

\author{Gr\'egoire Pihan}
\affiliation{Physics Department, University of Houston, Box 351550, Houston, TX 77204, USA}

\author{Bj\"orn Schenke}
\affiliation{Physics Department, Brookhaven National Laboratory, Upton, NY 11973, USA}

\author{Chun Shen}
\affiliation{Department of Physics and Astronomy, Wayne State University, Detroit, Michigan, USA}

\date{\today}

\begin{abstract}
Dissipative processes are pivotal for understanding the hydrodynamic evolution of hot and dense QCD matter created in relativistic nuclear collisions. The interplay of multiple conserved charges -- net baryon, strangeness, and electric charge -- is of particular interest. We simulate the longitudinal hydrodynamic evolution with the three diffusion currents in a hydrodynamic model with a lattice-QCD-based equation of state, \textsc{neos-4d}, and estimate rapidity distributions including diffusive corrections to the phase-space distribution in the presence of multiple charges, which ensure charge conservation at particlization. We determine the response of particle yields at midrapidity to changes in the diagonal and off-diagonal conductivities. Inversely, we find that most components of the conductivity matrix can be constrained experimentally using identified particle multiplicities at different collision energies. 
\end{abstract}


\maketitle

\section{Introduction}
\label{sec1}
\vspace*{-2mm}

Relativistic nuclear collisions are a powerful tool for exploring the phase diagram of quantum chromodynamics (QCD) \cite{Cabibbo:1975ig,Baym_diagram}. The experimental discovery of the quark-gluon plasma (QGP), a deconfined phase of QCD, at the Relativistic Heavy Ion Collider (RHIC) \cite{Adcox:2004mh,Adams:2005dq,Back:2004je,Arsene:2004fa,Harris:2023tti} has marked a milestone in particle and nuclear physics as a successful step toward mapping the phase diagram. The produced hot medium is quantified as a relativistic fluid with extremely small viscosity, allowing us to constrain the fundamental properties of QCD, such as the equation of state and transport coefficients, from the observation of produced particles. The Large Hadron Collider (LHC) \cite{Aamodt:2010pa,ATLAS:2011ah,Chatrchyan:2012wg} has confirmed that the fluidity persists at much higher temperatures, establishing that the hydrodynamic model serves as a standard framework of relativistic nuclear collisions \cite{Heinz:2024jwu}.

The QCD phase structure at vanishing (net-charge) densities is well constrained from lattice QCD, which finds a crossover transition \cite{Borsanyi:2010bp, Bazavov:2011nk, Borsanyi:2024xrx}. At high net baryon densities, on the other hand, details of the transition remain largely unknown, partly owing to the theoretical challenge in the lattice QCD simulations at finite chemical potentials known as the sign problem \cite{deForcrand:2010ys}. Model studies predict a variety of non-trivial phase structures, including a first-order phase transition between quark and hadronic phases in the dense regime and the associated critical end point \cite{Asakawa:1989bq}. One of the main objectives of the beam energy scan (BES) programs at RHIC \cite{STAR:2010vob} is to experimentally elucidate the phase structure and understand the macroscopic properties of dense quark matter. The system is known to be reasonably well described within hydrodynamic models down to $\sqrt{s_\mathrm{NN}}=\mathcal{O}(10)$ GeV. Generally, hydrodynamic frameworks are expected to be applicable to the study of dense QCD matter explored in future collider programs, including the GSI Facility for Antiproton and Ion Research (FAIR), JINR Nuclotron-based Ion Collider fAcility (NICA), IMP High Intensity heavy-ion Accelerator Facility (HIAF), and JAEA/KEK Japan Proton Accelerator Research Complex (J-PARC).

In addition to net baryon number (B), net electric charge (Q), and net strangeness (S) are also conserved during the spacetime evolution of the QCD medium. It is thus pivotal to consider a four-dimensional phase diagram for a quantitative understanding of relativistic nuclear collisions: The dynamics of multiple charges, supplemented by a four-dimensional QCD equation of state \cite{Werner:2010aa,Monnai:2019hkn,Noronha-Hostler:2019ayj,Aryal:2020ocm,Karthein:2021nxe,Schafer:2021csj,Monnai:2021kgu,Monnai:2024pvy,Plumberg:2024leb,Abuali:2025tbd,Pala:2025qoa}, is important for quantifying the dense quark matter created in nuclear collisions at the BES energies \cite{Roch:2025pcj}. It is also indispensable in the description of nuclear collisions at higher energies when interested in a wide rapidity range, for example when studying charge-dependent nuclear structure \cite{Pihan:2024lxw,Pihan:2025pep}.

Although the quark matter exhibits nearly-perfect fluidity, dissipative currents are finite \cite{Danielewicz:1984ww,Kovtun:2004de} and are known to play an essential role in quantitative predictions of transverse momentum and rapidity dependencies of particle distributions observed in nuclear collisions. Shear and bulk viscosities are associated with flow gradients that correspond to deformation and expansion/contraction of the medium, respectively. Baryon, charge, and strangeness diffusion, on the other hand, occur in finite-density systems as responses to fugacity gradients. Estimations of the conductivities have been a topic of interest \cite{Rougemont:2017tlu,Greif:2017byw,Fotakis:2019nbq,Das:2021bkz,Fotakis:2024hmz,Zhang:2024htn,Dey:2024hhc}. The effects of baryon diffusion have been studied extensively using the hydrodynamic model in, for example, Refs.~\cite{Monnai:2012jc,Denicol:2018wdp,Li:2018fow,Fotakis:2019nbq,Monnai:2019jkc,Du:2019obx,Wu:2021fjf,Pala:2025qoa}. The interplay of baryon, electric charge, and strangeness diffusion has also gained attention in the community \cite{Fotakis:2019nbq,Pala:2025qoa}.

In this manuscript, we investigate the effects of multiple diffusion currents in the QCD matter created in relativistic nuclear collisions to understand the relation between rapidity distributions and conductivities. First, we develop a consistent framework of the causal relativistic diffusive hydrodynamic model. The off-equilibrium corrections to the phase-space distribution, which are required for charge conservation at particlization, are computed based on Grad's moment method \cite{Grad:1949zza,Israel:1979wp}. We then estimate the effects of baryon, electric charge, and strangeness diffusion on the dynamical evolution of the charges with and without cross-coupling currents at different collision energies using numerical simulations focusing on the longitudinal dynamics. Rapidity distributions and $p_T$ spectra of conserved charges, including the $\delta f$ corrections, are calculated at particlization. We also investigate the sensitivity of the particle yields at midrapidity to the conductivities based on the linear response matrix, the quadratic response tensor, and Gaussian process regression models.

This paper is organized as follows. We review and discuss second-order relativistic hydrodynamics with multiple diffusion currents in Sec.~\ref{sec:eom}. Section~\ref{sec:df} is devoted to the derivation of the distribution function with diffusive corrections. Hydrodynamic simulations of conserved charges in nuclear collisions are presented in Sec.~\ref{sec:num}. We construct the response surfaces of multiplicities and conductivities as a method to constrain the transport coefficients from experimental data in Sec.~\ref{sec:sensitivity}. Section~\ref{sec:summary} presents a summary and conclusions. We use natural units $c=\hbar=k_B=1$ and the mostly-minus metric $g^{\mu \nu} = \mathrm{diag(+,-,-,-)}$ in this study. 

\section{Equations of motion}
\label{sec:eom}
\vspace*{-2mm}

We consider relativistic hydrodynamic evolution with diffusion processes of multiple conserved charges: baryon, electric charge, and strangeness. The energy-momentum tensor and conserved currents can be decomposed as 
\begin{eqnarray}
T^{\mu \nu} &=& (e + P + \Pi) u^\mu u^\nu - (P + \Pi) g^{\mu \nu} + \pi^{\mu \nu},\\
N_B^{\mu } &=& n_B u^\mu + V_B^{\mu },\\
N_Q^{\mu } &=& n_Q u^\mu + V_Q^{\mu },\\
N_S^{\mu } &=& n_S u^\mu + V_S^{\mu },
\end{eqnarray}
where $e$ is the energy density, $P$ the hydrostatic pressure, and $u^\mu$ the fluid four-velocity.
We choose the Landau frame with fluid velocities following energy-momentum currents, $T^{\mu\nu} u_\nu = e u^{\mu}$. In this frame, the heat current vanishes.
$n_B, n_Q$, and $n_S$ are the densities of net baryon, electric charge, and strangeness, and $V_B^\mu, V_Q^\mu$, and $V_S^\mu$ are the corresponding diffusion currents, respectively. The shear stress tensor $\pi^{\mu \nu}$ and the bulk pressure $\Pi$ will be dropped in this study for simplicity. 

The equations of motion are provided by the conservation laws
\begin{eqnarray}
\partial_\mu T^{\mu \nu} &=& 0,\\
\partial_\mu N_B^{\mu } &=& 0,\\
\partial_\mu N_Q^{\mu } &=& 0,\\
\partial_\mu N_S^{\mu } &=& 0,
\end{eqnarray}
and the constitutive relations.
A multi-component version of the second-order hydrodynamics with multiple conserved charges can be found, for instance, in Refs.~\cite{Monnai:2010qp,Kikuchi:2015swa,Fotakis:2022usk,Hu:2022vph,Almaalol:2022pjc}. 
Here, we consider the simplest causal and stable diffusion equations, obtained by retaining only the relaxation terms,
\begin{eqnarray}
 \begin{pmatrix}
 V_B^\mu \\ V_Q^\mu \\ V_S^\mu
  \end{pmatrix}&=& \mathcal{K}
 \begin{pmatrix}
 \nabla^\mu \frac{\mu_B}{T} \\ \nabla^\mu \frac{\mu_Q}{T} \\ \nabla^\mu \frac{\mu_S}{T}
  \end{pmatrix}
 - \mathcal{T} \Delta^{\mu}_{\ \nu} \begin{pmatrix}
 D V_B^\nu \\ D V_Q^\nu \\ D V_S^\nu
  \end{pmatrix}.\label{eq:eq}
\end{eqnarray}
$D=u^\mu\partial_\mu$ and $\Delta^{\mu \nu} \partial_\nu=\nabla^\mu$ where $\Delta^{\mu \nu} = g^{\mu \nu}-u^\mu u^\nu$ is a spatial projection operator. $\mathcal{K}$ and $\mathcal{T}$ are the transport coefficient matrices of conductivities
\begin{eqnarray}
 \mathcal{K} = \begin{pmatrix}
 \kappa_{BB} & \kappa_{BQ} & \kappa_{BS} \\
 \kappa_{QB} & \kappa_{QQ} & \kappa_{QS} \\
 \kappa_{SB} & \kappa_{SQ} & \kappa_{SS} 
  \end{pmatrix},
\end{eqnarray}
and of relaxation times
\begin{eqnarray}
 \mathcal{T} = \begin{pmatrix}
 \tau_{BB} & \tau_{BQ} & \tau_{BS} \\
 \tau_{QB} & \tau_{QQ} & \tau_{QS} \\
 \tau_{SB} & \tau_{SQ} & \tau_{SS} 
  \end{pmatrix}.
\end{eqnarray}
$\mathcal{K}$ is a positive semi-definite matrix because the entropy production must be positive. Also, it is a symmetric matrix owing to the Onsager reciprocal relations \cite{Onsager:1931jfa,Onsager:1931kxm}. $\mathcal{T}$, on the other hand, is not guaranteed to be symmetric. In this study, we assume that it is proportional to the identity matrix $\mathcal{I}$ as $\mathcal{T} = \tau_R \mathcal{I}$ to simplify the numerical simulations.

\section{Distortion of equilibrium distributions}
\label{sec:df}
\vspace*{-2mm}

\begin{figure*}[tb]
\includegraphics[width=7.2in]{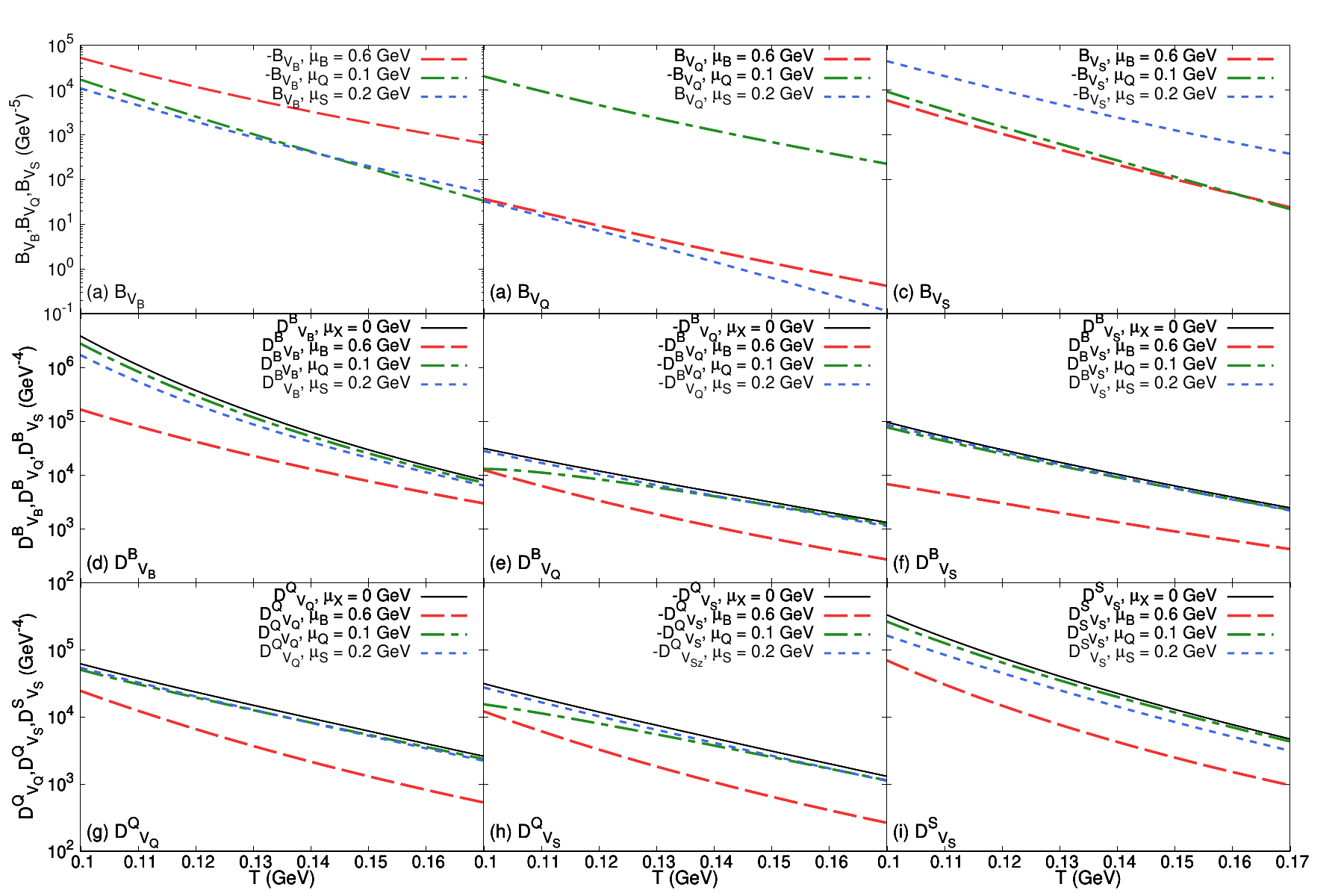}
\caption{The coefficients in the deformation tensors, (a) $\mathcal{B}_{V_B}$, (b) $\mathcal{B}_{V_Q}$, (c) $\mathcal{B}_{V_S}$, (d) $\mathcal{D}^B_{V_B}$, (e) $\mathcal{D}^B_{V_Q}(=\mathcal{D}^Q_{V_B})$, (f) $\mathcal{D}^B_{V_S}(=\mathcal{D}^S_{V_B})$, (g) $\mathcal{D}^Q_{V_Q}$, (h) $\mathcal{D}^Q_{V_S}(=\mathcal{D}^S_{V_Q})$, and (i) $\mathcal{D}^S_{V_S}$ at $\mu_B = \mu_Q = \mu_S = 0$ GeV (solid line), $\mu_B = 0.6$ GeV and $\mu_Q = \mu_S = 0$ GeV (dashed line), $\mu_Q = 0.1$ GeV and $\mu_B = \mu_S = 0$ GeV (dashed-dotted line), and $\mu_S = 0.2$ GeV and $\mu_B = \mu_Q = 0$ GeV (dotted line).}
\label{fig:coeff}
\end{figure*}

The phase-space distributions deform from their equilibrium form in the presence of diffusion currents. Here they are separated into the equilibrium and off-equilibrium components as
\begin{eqnarray}
f^i = f_0^i + \delta f^i ,
\end{eqnarray}
where the index $i$ denotes particle species. Bose-Einstein or Fermi-Dirac statistics determines $f_0^i = \{\exp[(E_i-\mu_i)/T] \mp 1\}^{-1}$, where $E_i = \sqrt{\mathbf{p}^2 + m_i^2}$ is the energy, $m_i$ is the mass, and $\mathbf{p}^2$ is the three-momentum squared. The chemical potential is defined as $\mu_i = B_i \mu_B + Q_i \mu_Q + S_i \mu_S$ where $B_i$, $Q_i$, and $S_i$ are the quantum numbers of net baryon, electric charge, and strangeness, respectively. The upper sign is for bosons, and the lower sign is for fermions. 

Assuming that the deformation of the distribution is expressed in terms of the diffusion currents, Grad's moment method \cite{Israel:1979wp} can be extended to a system with multiple conserved charges \cite{Monnai:2010qp} as 
\begin{eqnarray}
\label{eq:df}
\delta f^i &=& - f_0^i (1\mp f_0^i) ( \sum_X X_i p_i^\mu \varepsilon_{\mu}^X
+ p_i^\mu p_i^\nu \varepsilon_{\mu \nu} ) ,
\end{eqnarray}
where the deformation tensors are 
\begin{eqnarray}
\varepsilon_{\mu}^X &=& \sum_{Y}  \mathcal{D}_{V_Y}^X V_\mu^Y, \\
\varepsilon_{\mu \nu} &=& \sum_{X} \mathcal{B}_{V_X} (V_\mu^X u_\nu + u_\mu V_\nu^X) .
\end{eqnarray}
We denote arbitrary charges with symbols $X, Y, Z \in \{B,Q,S\}$.

It is straightforward to implement the shear and bulk viscous $\delta f$ \cite{Teaney:2003kp,Monnai:2009ad,Denicol:2009am,Roch:2025pcj} should it become necessary.
Note that the shear and bulk viscous corrections are separated into the scalar and tensor moments in $\delta f$, which do not interfere with the charge diffusion modes.

The coefficients $\mathcal{D}$'s and $\mathcal{B}$'s are determined from the self-consistency conditions
\begin{eqnarray}
\label{eq:Vx}
V_X^\mu &=& \Delta^{\mu}_{\ \nu} \sum_i \int \frac{X_i g_i d^3 p}{(2\pi)^3 E_i} p^\nu \delta f_i,
\end{eqnarray}
in kinetic theory. Here $E_i$ is the energy and $g_i$ is the degeneracy. Further, because the energy dissipation vanishes in the Landau frame, one has\begin{eqnarray}
\label{eq:W}
0 &=& \Delta^{\mu}_{\ \nu} u_\alpha \sum_i \int \frac{g_i d^3 p}{(2\pi)^3 E_i} p^\nu p^\alpha \delta f_i.
\end{eqnarray}
Equations~(\ref{eq:Vx}) and (\ref{eq:W}) can be expressed as
\begin{eqnarray}
\begin{pmatrix}
0 \\
V_B^\mu \\
V_Q^\mu \\
V_S^\mu 
\end{pmatrix} \hspace{-0.1cm}= \hspace{-0.1cm} - \hspace{-0.1cm}
\begin{pmatrix}
2 J_{41} & J^B_{31} & J^Q_{31} & J^S_{31} \\
2 J^B_{31} & J^{BB}_{21} & J^{BQ}_{21} & J^{BS}_{21} \\
2 J^Q_{31} & J^{BQ}_{21} & J^{QQ}_{21} & J^{QS}_{21} \\
2 J^S_{31} & J^{BS}_{21} & J^{QS}_{21} & J^{SS}_{21} 
\end{pmatrix} \hspace{-0.2cm}
\begin{pmatrix}
\Delta^{\mu}_{\ \nu} u_\alpha\varepsilon^{\nu \alpha} \\
\Delta^{\mu}_{\ \nu} \varepsilon^{\nu}_B \\
\Delta^{\mu}_{\ \nu} \varepsilon^{\nu}_Q \\
\Delta^{\mu}_{\ \nu} \varepsilon^{\nu}_S 
\end{pmatrix}
\end{eqnarray}
using the moments defined as
\begin{eqnarray}
J^{XY\cdots}_{k l} &=& \frac{1}{(2l+1)!!} \sum_i \int \frac{(X_i Y_i\cdots) g_i d^3p}{(2\pi)^3 E_i} \nonumber \\
&\times& E_i^{k-2l} (- \mathbf{p}^2)^{l}  f_0^i(1\mp f_0^i) \,,
\end{eqnarray}
where the index ${XY\cdots}$ denotes additional weights of quantum numbers $(X_i Y_i\cdots)$ in the summation over $i$. 

The equations can be solved explicitly and the coefficients are expressed as 
\begin{eqnarray}
\mathcal{B}_{V_X} &=& (J^{X}_{31} J^{YY}_{21} J^{ZZ}_{21} + J^{Y}_{31} J^{XZ}_{21} J^{YZ}_{21} \nonumber \\
&+& J^{Z}_{31} J^{XY}_{21} J^{YZ}_{21} - J^{X}_{31} {J^{YZ}_{21}}^2 \nonumber \\ &-& J^{Y}_{31} J^{XY}_{21} J^{ZZ}_{21} - J^{Z}_{31} J^{XZ}_{21} J^{YY}_{21} )/ \mathcal{J}_4 ,\\
\mathcal{D}^{X}_{V_X} &=& (2 J_{41} {J^{YZ}_{21}}^2 + 2 {J^{Y}_{31}}^2 J^{ZZ}_{21} + 2 {J^{Z}_{31}}^2 J^{YY}_{21}  \nonumber \\ 
&-& 2 J_{41} J^{YY}_{21} J^{ZZ}_{21} - 4 J^{Y}_{31} J^{YZ}_{21} J^{Z}_{31})/ \mathcal{J}_4 ,\\
\mathcal{D}^{X}_{V_Y} &=& \mathcal{D}^{Y}_{V_X} = (2 J_{41} J^{XY}_{21} J^{ZZ}_{21} + 2 J^{X}_{31} J^{Z}_{31} J^{YZ}_{21} \nonumber \\ 
&+& 2 J^{Y}_{31} J^{Z}_{31} J^{XZ}_{21} -2 J_{41} J^{XZ}_{21} J^{YZ}_{21}  \nonumber \\ &-& 2 J^{X}_{31} J^{Y}_{31} J^{ZZ}_{21} -
2 {J^{Z}_{31}}^2 J^{XY}_{21} )/ \mathcal{J}_4 ,
\end{eqnarray}
where $X\neq Y\neq Z$. 
The denominator reads
\begin{eqnarray}
\mathcal{J}_4 &=& 2J_{41}(J^{BB}_{21}J^{QQ}_{21}J^{SS}_{21}+2J^{BQ}_{21}J^{BS}_{21}J^{QS}_{21} 
\nonumber \\
&-& 
 J^{BB}_{21} {J^{QS}_{21}}^2 -  J^{QQ}_{21} {J^{BS}_{21}}^2 - J^{SS}_{21} {J^{BQ}_{21}}^2 )
\nonumber \\
&+& 2{J^{B}_{31}}^2 ({J^{QS}_{21}}^2 - J^{QQ}_{21}J^{SS}_{21}) 
\nonumber \\
&+&
2{J^{Q}_{31}}^2 ({J^{BS}_{21}}^2  - J^{BB}_{21}J^{SS}_{21})
\nonumber \\
&+& 2{J^{S}_{31}}^2 ({J^{BQ}_{21}}^2  - J^{BB}_{21}J^{QQ}_{21} ) 
\nonumber \\
&+& 4J^{B}_{31}J^{Q}_{31}(J^{BQ}_{21}J^{SS}_{21} - J^{BS}_{21}J^{QS}_{21})
\nonumber \\
&+& 4J^{B}_{31}J^{S}_{31} (J^{BS}_{21}J^{QQ}_{21} - J^{BQ}_{21}J^{QS}_{21})
\nonumber \\
&+&  4J^{Q}_{31}J^{S}_{31}(J^{BB}_{21}J^{QS}_{21} - J^{BQ}_{21}J^{BS}_{21})
\label{eq:detJ4} .
\end{eqnarray}

The coefficients are plotted at different chemical potentials as functions of temperature in Fig.~\ref{fig:coeff}. Note that they can be negative depending on chemical potentials and some of the results are flipped in sign as indicated in the legend. $\mathcal{B}_{V_{X}}$ are zero in the limit of vanishing chemical potentials because they are anti-symmetric under the simultaneous inversion of chemical potentials. They are also most sensitive to the respective chemical potentials $\mu_X$. The magnitudes of the diagonal $\mathcal{D}^X_{V_{X}}$ tend to be larger than those of the off-diagonal $\mathcal{D}^X_{V_{Y}}$. They remain finite at zero chemical potentials, unlike $\mathcal{B}_{V_{X}}$. 

\section{Numerical simulations}
\label{sec:num}
\vspace*{-2mm}

We demonstrate the diffusion processes in relativistic nuclear collisions, focusing on longitudinal dynamics, by constructing a (1+1)-dimensional numerical hydrodynamic model. The hydrodynamic evolution of net baryon, electric charge and strangeness is estimated using a lattice QCD based EoS model, \textsc{neos-4d} \cite{Monnai:2024pvy}, which introduces an efficient parameterization method that renders the four-dimensional relation of thermodynamic variables \cite{Monnai:2019hkn} directly usable in hydrodynamic simulations.

The initial conditions are parametrized as
\begin{eqnarray}
e(\eta_s) &=& e_0 \exp(-a_1 \eta_s^2 - a_2 \eta_s^4), \\
n_B(\eta_s) &=& n_B^+(\eta_s) + n_B^-(\eta_s),\\
n_Q(\eta_s) &=& c_0 n_B(\tau_0,\eta_s),\\
n_S(\eta_s) &=& 0,
\end{eqnarray}
where
\begin{align}
&n_B^\pm(\eta_s)  \nonumber \\
&=\begin{cases}
n_0 \exp[-b_1 (\eta_s \mp \eta_0)^2 - b_2 (\eta_s \mp \eta_0)^4] &   (|\eta_s| > \eta_0), \\
\displaystyle n_0 \bigg\{(1-b_4) \exp[-b_3 (\eta_s \mp \eta_0)^2] + \frac{b_4}{2} \bigg\}&  (|\eta_s| \leq \eta_0).
\end{cases}\nonumber \\
\end{align}
The parameters for the shape $a_1, a_2, \eta_0, b_1, b_2, b_3$, and $b_4$ are tuned to approximately reproduce the PHOBOS data for 0-6\% Au+Au collisions at (i) $\sqrt{s_\mathrm{NN}} = 19.6$ GeV, (ii) 62.4 GeV, and (iii) 200 GeV \cite{Back:2002wb,PHOBOS:2005zhy,PHOBOS:2010eyu} (Table~\ref{table:initial}). The BRAHMS \cite{BRAHMS:2003wwg} and STAR BES \cite{STAR:2008med,STAR:2017sal} data are also used to constrain the net baryon distributions. $\tau_0$ is the initial proper time. $S$ is the transverse area of the system, which is assumed to be larger than the overlapping region of the two colliding nuclei to compensate for the lack of transverse expansion. $e_0$ is calibrated with charged particle pseudo-rapidity distributions from the PHOBOS collaboration. $n_0$ is constrained so that the integrated net baryon number matches the average number of participants calculated in the Monte-Carlo Glauber model. 
$c_0$ is set to 0.4 to reflect the charge-to-baryon ratio in heavy nuclei.

\begin{table}[tb]
{\begin{tabular}{c|c|c|c|c|c|c|c|c|c}
\hline \hline
  & $\tau_0$(fm/$c$) & $S$(fm$^2$)& $a_1$ & $a_2$ & $\eta_0$ & $b_1$ & $b_2$ & $b_3$ & $b_4$ \\ \hline
(i) & 3.0 & 200 & 0.41  & 0.021 & 1.0 & 0.77 & 3.2 & 3.2 & 0.37 \\ \hline
(ii) & 1.5 & 200 & 0.13 & 0.015 & 2.1 & 0.78 & 1.2 & 0.81 & 0.21 \\ \hline
(iii) & 1.0 & 250 & 0.02 & 0.0095 & 3.0 & 0.79 & 0.72 & 0.52 & 0.13 \\ \hline \hline
\end{tabular}
\caption{The parameters for the initial conditions for (i) $\sqrt{s_\mathrm{NN}} = 19.6$ GeV, (ii) 62.4 GeV, and (iii) 200 GeV.} 
\label{table:initial}}
\end{table}

For demonstrative purposes, the diagonal conductivities are set to $\kappa_{BB}/T^2= 0.05$, $\kappa_{QQ}/T^2= 0.02$, and $\kappa_{SS}/T^2= 0.1$, and the off-diagonal conductivities to $\kappa_{BQ}/T^2= 0.002$, $\kappa_{BS}/T^2= -0.02$, and $\kappa_{QS}/T^2= 0.01$ unless otherwise specified. This choice of parameters guarantees that the matrix of conductivities is positive semi-definite. Those values are of the same order as the ones proposed in holographic \cite{Rougemont:2017tlu} and kinetic \cite{Greif:2017byw} approaches. 
We will consider the two cases where the off-diagonal coefficients are turned on and off to evaluate the effects of the cross-coupling currents in this section. The relaxation time is chosen as $\tau_R = \ln2/(2\pi T)$ based on a holographic approach \cite{Natsuume:2007ty}. Fine-tuning of the coefficients is left for future studies.

\begin{figure*}[tb]
\includegraphics[width=7.2in]{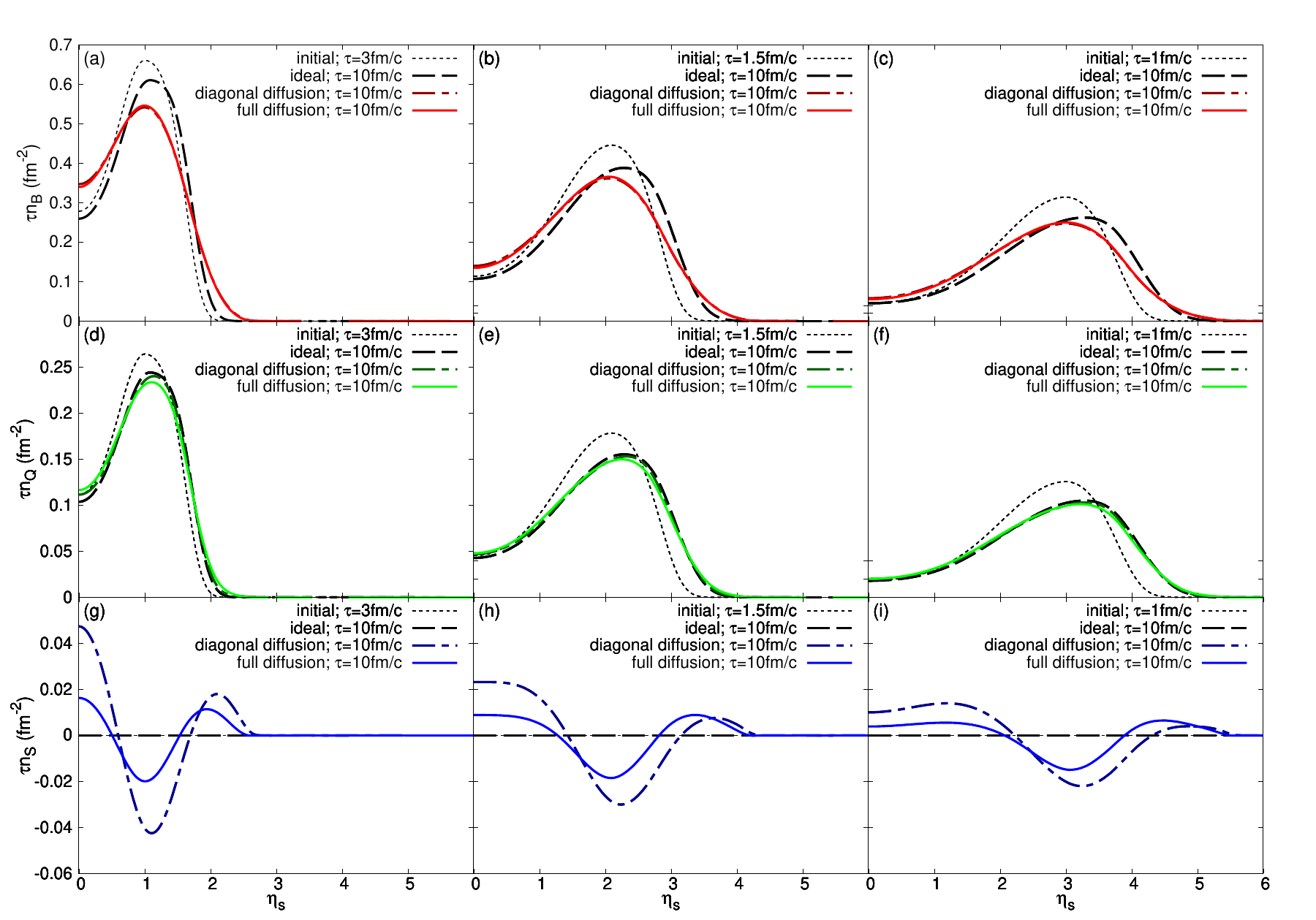}
\caption{
Spacetime rapidity distributions of net baryon density at the initial time (thin dotted line) and 10 fm/$c$ (thick lines) at (a) $\sqrt{s_\mathrm{NN}} = 19.6$ GeV, (b) $62.4$ GeV, and (c) $200$ GeV. The same for (d)-(f) net charge and (g)-(i) net strangeness densities, respectively. Dashed, dash-dotted, and solid lines denote ideal, diagonal diffusion, and full diffusion cases, respectively.
}
\label{fig:densities}
\end{figure*}

The Cooper-Frye prescription of the particlization \cite{Cooper:1974mv} is affected by off-equilibrium corrections \cite{Teaney:2003kp,Monnai:2009ad,Denicol:2009am}:
\begin{eqnarray}
E_i \frac{dN^i}{d^3 p} = \frac{g_i}{(2\pi)^3} \int_\Sigma p_i^\mu d\sigma_\mu (f_0^i + \delta f^i).\label{eq:particlization}
\end{eqnarray}
$\delta f^i$ is essential for the charge conservation at particlization. Its explicit form in the presence of multiple diffusion currents is as derived in Sec.~\ref{sec:df}. The particlization hypersurface $\Sigma$ is defined by $e=0.4$ GeV/fm$^3$. 

\subsection{Hydrodynamic evolution}
\label{sec:evo}

Figure \ref{fig:densities} shows the time evolution of net baryon, electric charge, and strangeness distributions in spacetime rapidity for 0-6\% central Au+Au collisions at $\sqrt{s_\mathrm{NN}} = 19.6$ GeV, 62.4 GeV, and 200 GeV. One can see that the flow induced by pressure gradients carries the densities to forward rapidities for the net baryon and electric charge distributions in the ideal hydrodynamic case. The convection effect becomes larger at higher collision energies. The strangeness distributions, on the other hand, stay at zero under the inviscid evolution. 

\begin{figure*}[tb]
\includegraphics[width=3.3in]{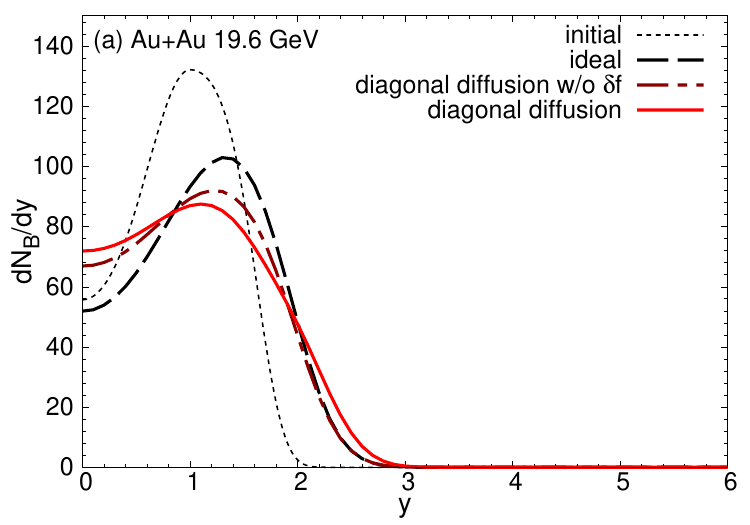}
\includegraphics[width=3.3in]{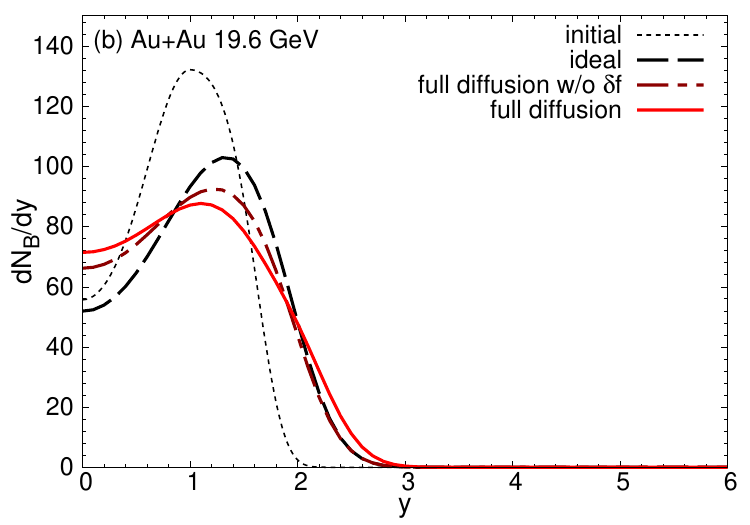}
\includegraphics[width=3.3in]{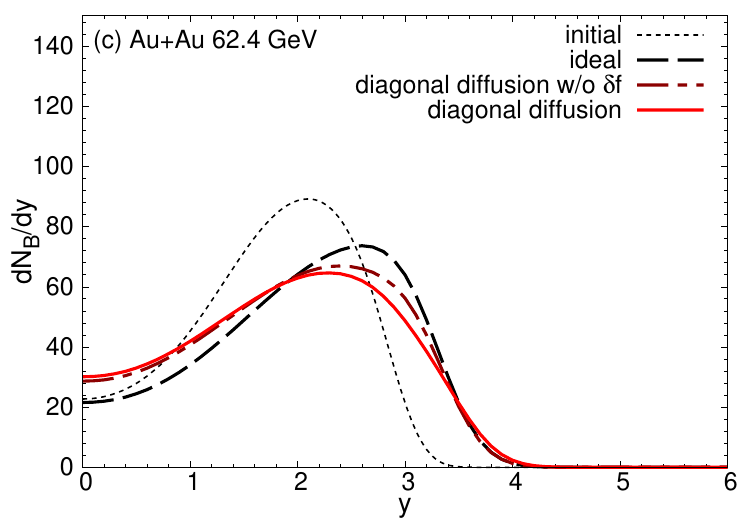}
\includegraphics[width=3.3in]{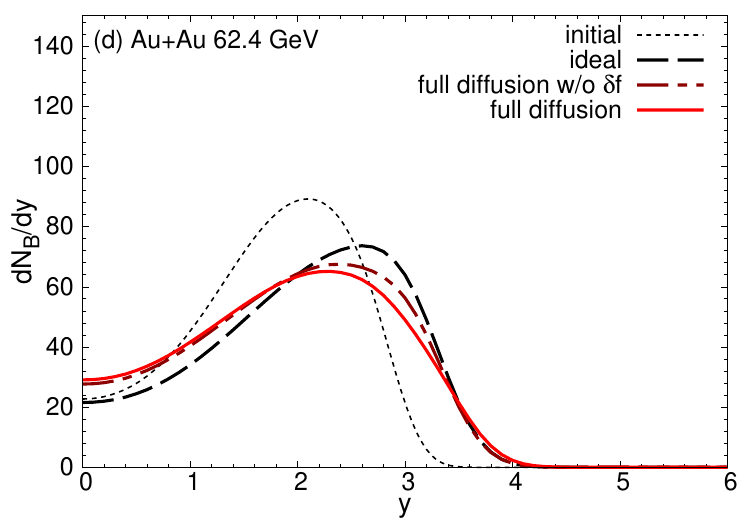}
\includegraphics[width=3.3in]{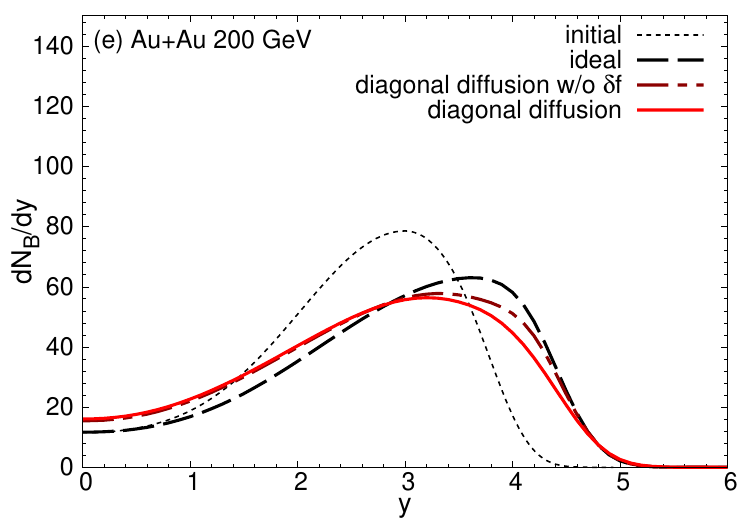}
\includegraphics[width=3.3in]{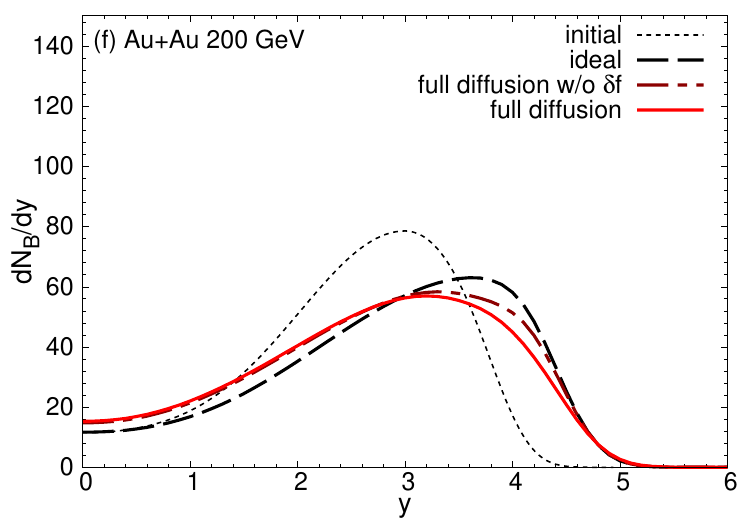}
\caption{Net baryon rapidity distributions for the (a) diagonal and (b) full diffusive cases, with and without $\delta f$ corrections (solid and dashed-dotted lines, respectively) compared to the ideal case (dashed lines) at $\sqrt{s_\mathrm{NN}} = 19.6$ GeV. The same at (c)(d) $\sqrt{s_\mathrm{NN}} = 62.4$ GeV and (e)(f) $200$ GeV. Initial net baryon distributions (dotted lines) are plotted for comparison assuming $\eta_s$ equals $y$.
}
\label{fig:dnbdy}
\end{figure*}

The effect of diffusion with diagonal terms is evident in the net baryon distribution, which is broadened in spacetime rapidity owing to the gradients in the fugacity (Fig. \ref{fig:densities} (a)-(c)) \cite{Monnai:2012jc,Denicol:2018wdp,Li:2018fow}. Baryon diffusion tends to reduce the mean spacetime rapidity of the distribution with the effect becoming larger at lower energies. The effects of off-diagonal diffusion are small with the current parameter choices. 

The net electric charge distributions in Fig.\,\ref{fig:densities} (d)-(f) show that electric charge diffusion has much smaller effects in the absence of the cross-coupling terms, because its diagonal thermodynamic force is much smaller in the system where $|\mu_Q|\ll|\mu_B|$. 
The small shift of electric charge towards mid-spacetime rapidity can be caused by gradients in $\mu_Q/T$. The inhomogeneous $n_Q/n_B$ ratio in the medium induced by baryon diffusion can dynamically modify the gradients.
The off-diagonal diffusion tends to further broaden the net charge distribution but has only marginal effects for the chosen $\kappa_{BQ}$.

The strangeness diffusion current is non-vanishing even in a system with strangeness-free initial conditions, because the strangeness neutrality condition $n_S=0$ implies $\mu_S>0$ when $\mu_B>0$. As a result, strangeness is induced locally, producing a characteristic wave-like structure \cite{Fotakis:2019nbq} in the spacetime rapidity distribution (Fig.\,\ref{fig:densities} (g)-(i)). The effect of strangeness diffusion is found to become more pronounced at lower energies.
The cross-coupling currents strongly affect the net strangeness distribution and reduce the amplitude of the structure for the chosen negative $\kappa_{BS}$.

\begin{figure*}[tb]
\includegraphics[width=3.3in]{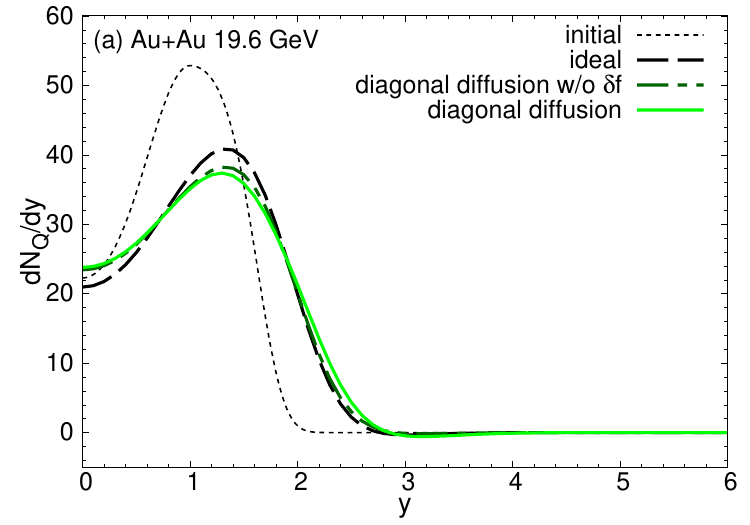}
\includegraphics[width=3.3in]{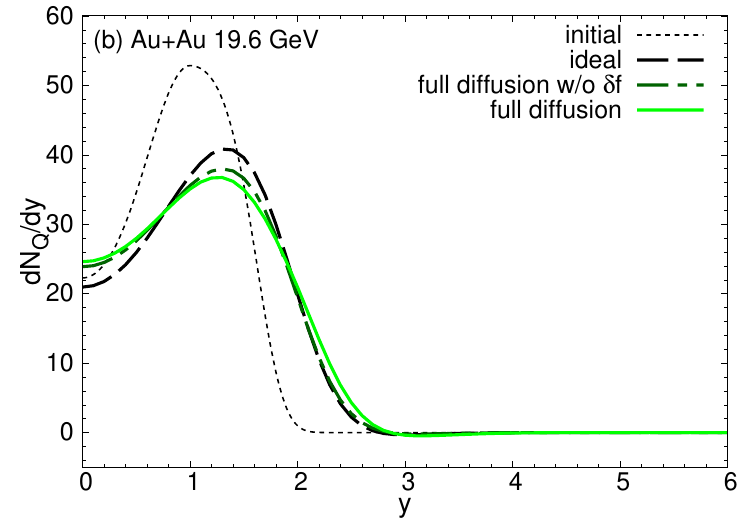}
\includegraphics[width=3.3in]{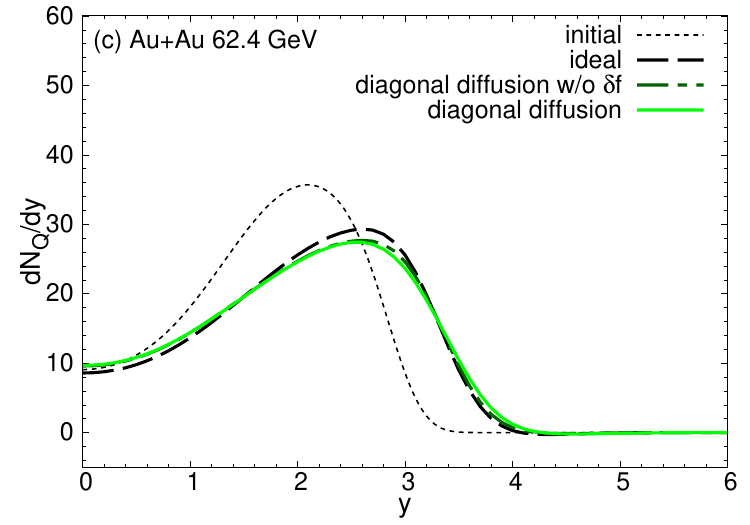}
\includegraphics[width=3.3in]{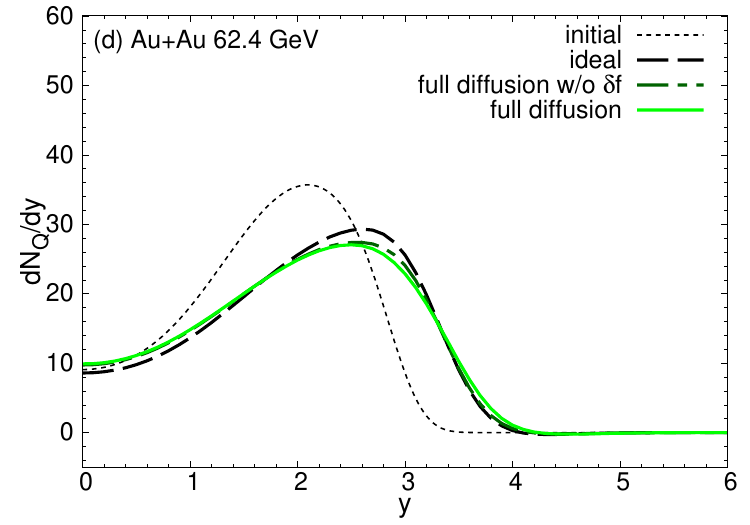}
\includegraphics[width=3.3in]{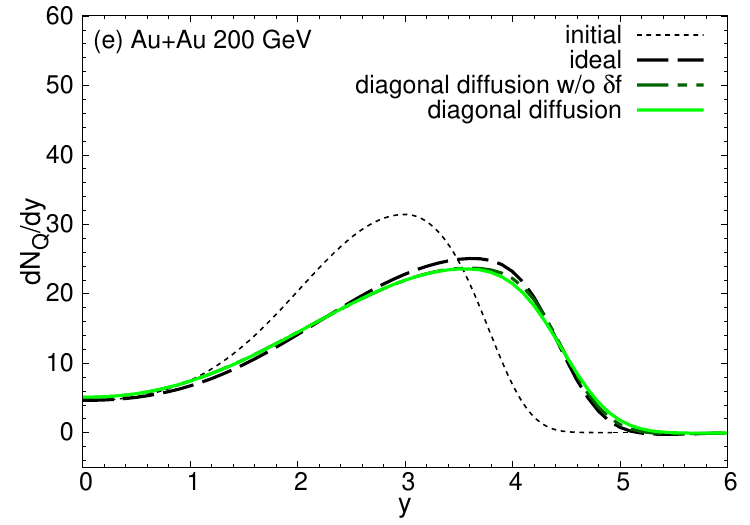}
\includegraphics[width=3.3in]{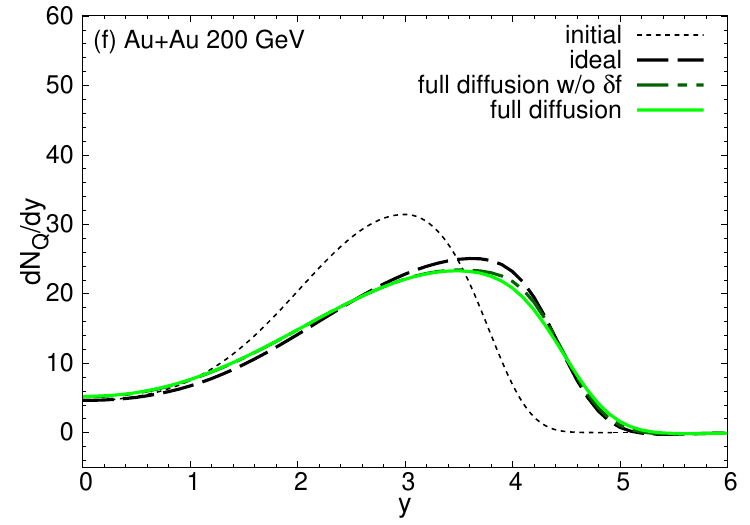}
\caption{Net electric charge rapidity distributions for the (a) diagonal and (b) full diffusive cases, with and without $\delta f$ corrections (solid and dashed-dotted lines, respectively) compared to the ideal case (dashed lines) at $\sqrt{s_\mathrm{NN}} = 19.6$ GeV. The same at (c)(d) $\sqrt{s_\mathrm{NN}} = 62.4$ GeV and (e)(f) $200$ GeV. Initial net electric charge distributions (dotted lines) are plotted for comparison assuming $\eta_s$ equals $y$.}
\label{fig:dnqdy}
\end{figure*}

\begin{figure*}[tb]
\includegraphics[width=3.3in]{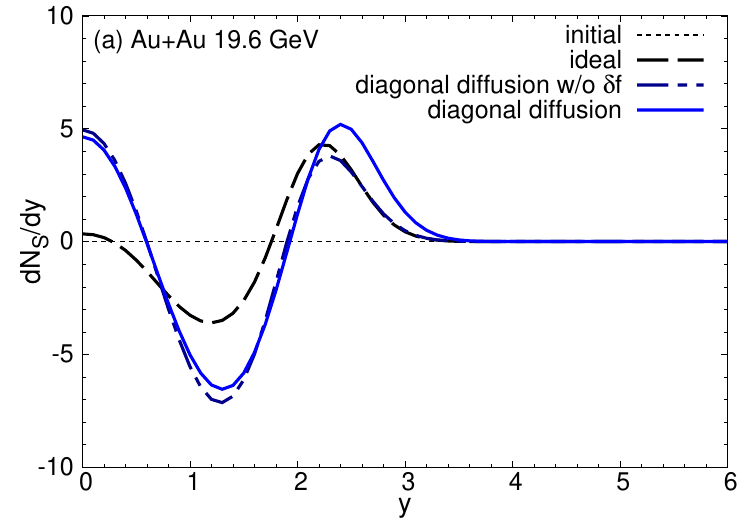}
\includegraphics[width=3.3in]{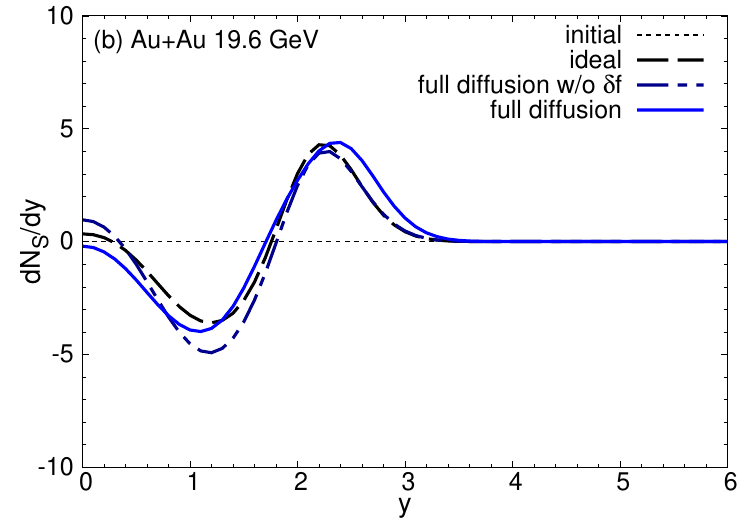}
\includegraphics[width=3.3in]{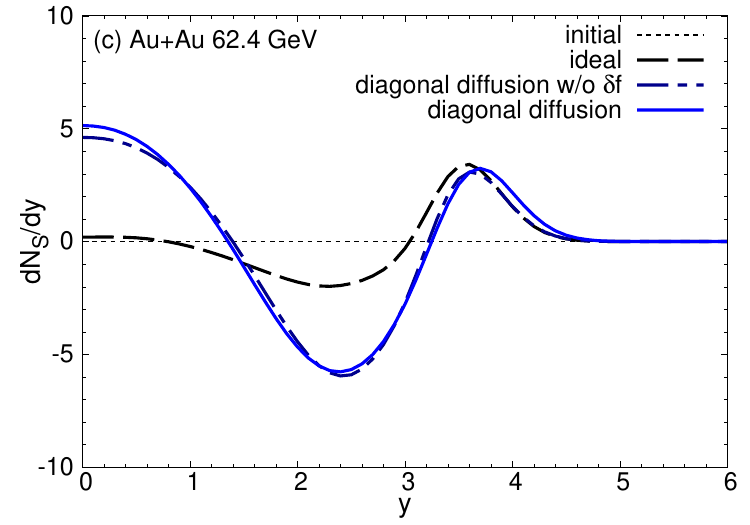}
\includegraphics[width=3.3in]{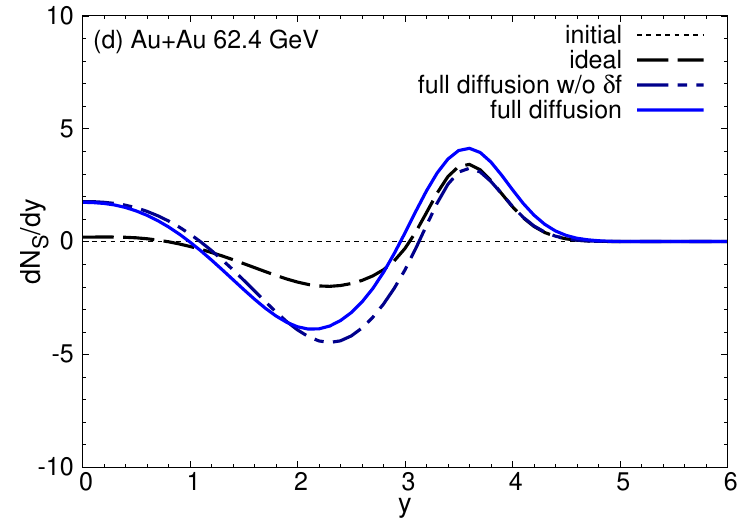}
\includegraphics[width=3.3in]{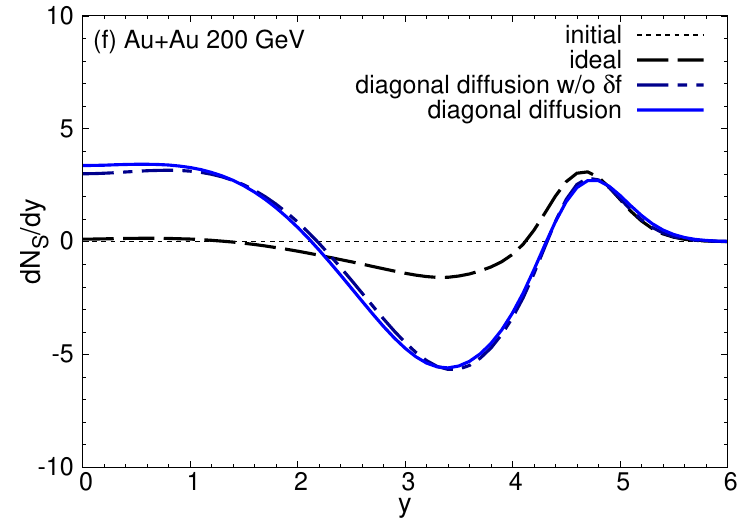}
\includegraphics[width=3.3in]{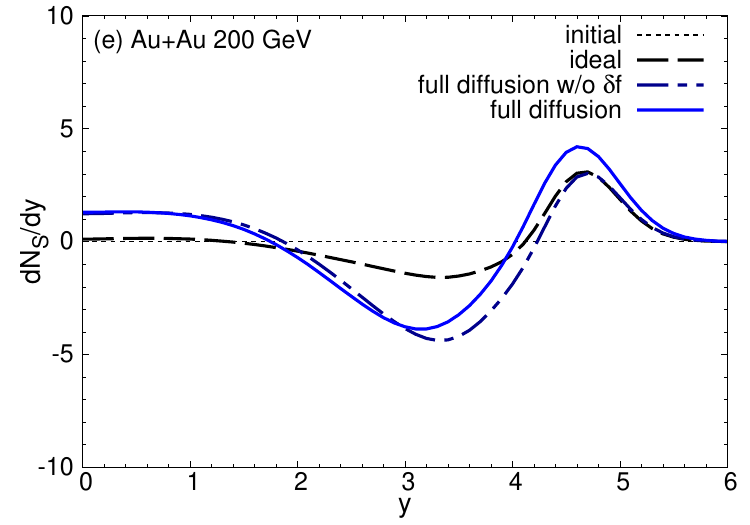}
\caption{Net strangeness rapidity distributions for the (a) diagonal and (b) full diffusive cases, with and without $\delta f$ corrections (solid and dashed-dotted lines, respectively) compared to the ideal case (dashed lines) at $\sqrt{s_\mathrm{NN}} = 19.6$ GeV. The same at (c)(d) $\sqrt{s_\mathrm{NN}} = 62.4$ GeV and (e)(f) $200$ GeV. Initial net strangeness distributions (dotted lines) are plotted for comparison assuming $\eta_s$ equals $y$.}
\label{fig:dnsdy}
\end{figure*}

\subsection{Rapidity distributions}
\label{sec:rap_dist}

We next estimate the rapidity distributions of conserved charges at particlization and elucidate the effects of diffusion from the modification of spacetime evolution and the distortion of the phase-space distribution $\delta f$ in the flow-to-particle conversion process. 

Figure~\ref{fig:dnbdy} shows the rapidity distributions of net baryons from ideal and diffusive hydrodynamic evolution with and without the $\delta f$ corrections at different energies. Baryon diffusion leads to a shift of baryon density towards midrapidity \cite{Monnai:2012jc,Denicol:2018wdp,Monnai:2019jkc}, and its effect becomes monotonically larger from $\sqrt{s_\mathrm{NN}}=200$ GeV to 62.4 GeV to 19.6 GeV, which is consistent with the results of hydrodynamic evolution observed in Sec.~\ref{sec:evo}. The effects of cross-coupling currents are mostly negligible at all collision energies. The $\delta f$ correction visibly modifies the shape of the distribution, and leads to an approximately 1-2\% correction to the integrated baryon number at $\sqrt{s_\mathrm{NN}}=19.6$~GeV and a 2-3\% correction at $\sqrt{s_\mathrm{NN}}=200$~GeV. $dN_B/dy$ is noticeably modified at midrapidity by $\delta f$ in the 19.6 GeV case.

The rapidity distributions of net electric charge are shown in Fig.~\ref{fig:dnqdy}. Diffusion processes of net electric charge are found to have a smaller impact than those of net baryon charge over the entire considered range of collision energies, which is consistent with the spacetime evolution in Fig.~\ref{fig:densities}. The broadening is in part caused by modifications of the bulk medium profiles owing to baryon diffusion. Comparing the results with diagonal diffusion and those with full diffusion, one can see that off-diagonal diffusion is small, possibly because of the choice of $\kappa_{BQ}$. The $\delta f$ corrections have only limited effects. The distributions become slightly negative near $y=3$, 4, and 5 at $\sqrt{s_\mathrm{NN}} = 19.6$ GeV, 62.4 GeV, and 200 GeV, respectively, because of the differences in the thermal smearing width among different charged hadrons, which will be explained later in detail.

\begin{figure}[tb]
\includegraphics[width=3.2in]{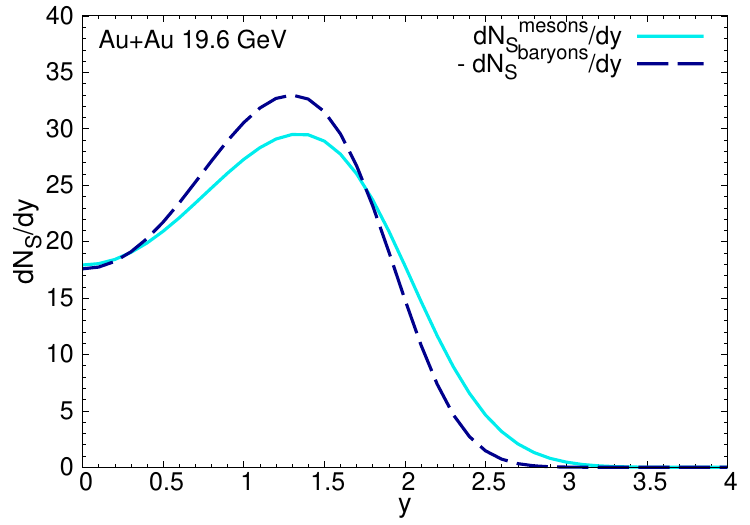}
\caption{Net strangeness rapidity distribution of mesons (solid line) compared with that of baryons with inverted sign (dashed line) from the ideal hydrodynamic model.}
\label{fig:dnsdy_mesons_baryons}
\end{figure}

The net strangeness rapidity distributions have characteristic wave-like structures (Fig.\,\ref{fig:dnsdy}). It is noteworthy that non-zero local strangeness can be observed even in the ideal hydrodynamic case \cite{Roch:2025pcj}. This is caused by the aforementioned thermal smearing in momentum space. Baryons with strangeness, such as $\Lambda$, have negative contributions to net strangeness because net baryon number is positive, \textit{i.e.}, there are more $s$ quarks than $\bar{s}$ quarks in the baryonic sector. On the other hand, mesons, such as kaons, have positive contributions to net strangeness because the strangeness chemical potential is positive. Since lighter particles have more thermal broadening in momentum space (see Fig.\,\ref{fig:dnsdy_mesons_baryons} for an example case at $\sqrt{s_\mathrm{NN}}=19.6$ GeV without diffusion), the competition between the contributions from the meson and baryon sectors leaves the wave-like imprint on the net strangeness rapidity distributions. The amplitude of the structure is determined by the temperature on the particlization hypersurface and thus is not much dependent on the collision energy.

Diffusion has a large impact on $dN_S/dy$. The strangeness transport by diffusion currents amplifies the wave-like structure. The existence of the cross-coupling current tends to reduce the structure when the off-diagonal conductivity $\kappa_{BS}$ is negative, which is consistent with the results of spacetime evolution in Fig.~\ref{fig:densities}.

\begin{figure}[tb]
\includegraphics[width=3.2in]{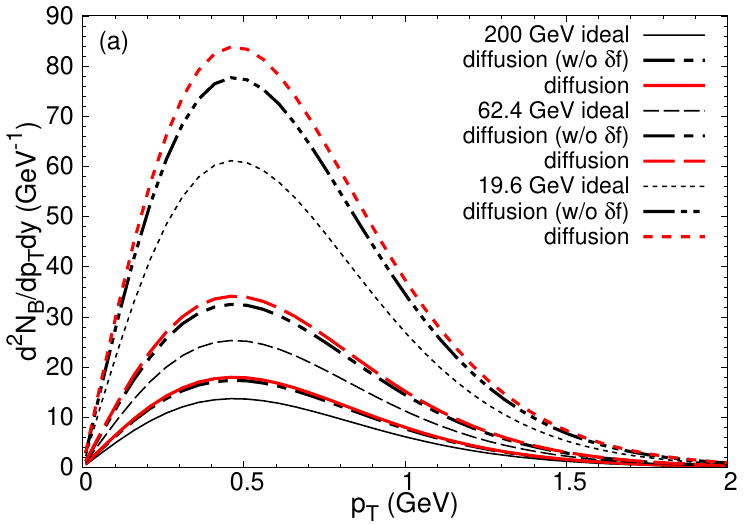}
\includegraphics[width=3.2in]{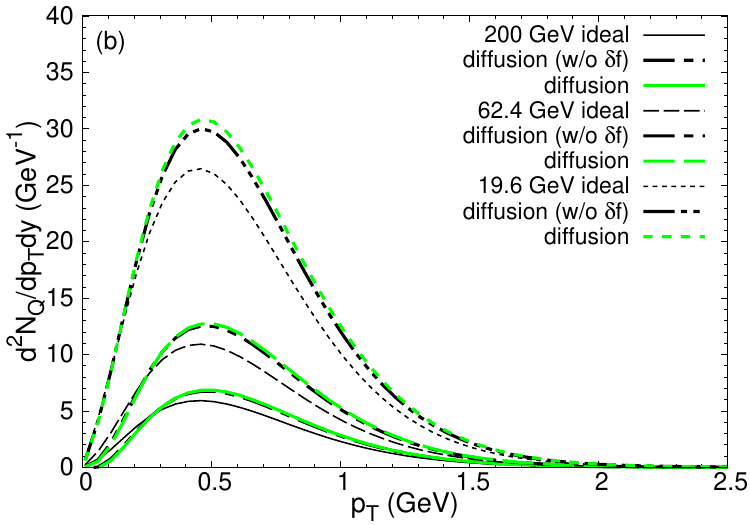}
\includegraphics[width=3.2in]{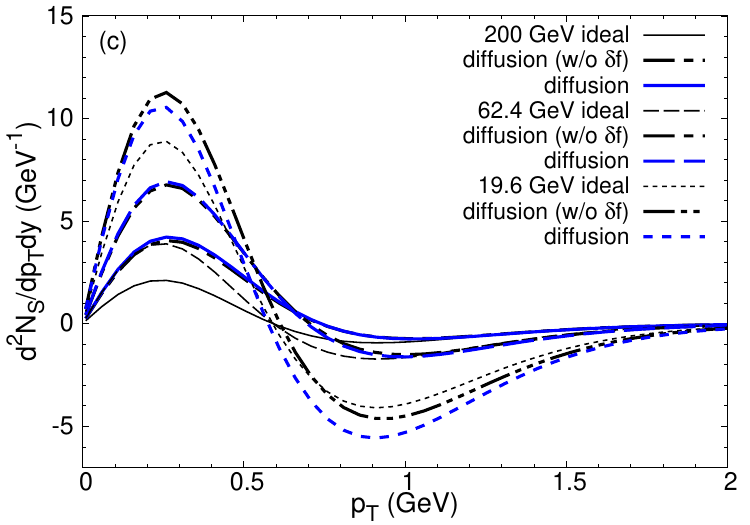}
\caption{$p_T$ spectra of (a) net baryon, (b) charge, and (c) strangeness. Thin solid, dashed, and dotted lines denote ideal hydrodynamic results at $\sqrt{s_\mathrm{NN}} = 200$ GeV, 62.4 GeV, and 19.6 GeV, respectively. The thick dash-dotted, dash-double-dotted, and dash-triple-dotted lines are the results with full diffusion without $\delta f$ corrections, and the thick solid, dashed, and dotted lines are those with $\delta f$ corrections.
}
\label{fig:dndptdy}
\end{figure}

\subsection{Transverse momentum spectra}
\label{sec:pt_sp}

We present the computed transverse momentum spectra of conserved charges $dN_X/dp_Tdy$ at midrapidity for the different collision energies in Fig.~\ref{fig:dndptdy}. The yields tend to become larger for lower collision energies and are enhanced in the presence of the diffusion currents, which is in line with the results of integrated $dN_X/dy$ in Figs.~\ref{fig:dnbdy}-\ref{fig:dnsdy}. The net strangeness spectra become negative because of the difference in the thermal smearing of mesons and baryons.

The $\delta f$ corrections of the diffusion currents on $p_T$ spectra are visible in baryons and strangeness at lower energies. Their $p_T$ dependence is found to be small, unlike that of shear and bulk viscosities \cite{Teaney:2003kp,Monnai:2009ad,Dusling:2011fd}. The bulk pressure $\Pi$ is a scalar, so the reduction of effective pressure $P+\Pi$ caused by the rapid longitudinal expansion of the medium also reduces the pressure in the transverse directions, leading to a smaller mean $p_T$. The shear stress tensor $\pi^{\mu\nu}$ is a traceless tensor, and suppression of $\pi^{zz}$ results in enhancement in $\pi^{xx}$ and $\pi^{yy}$ even for one-dimensional expansion, leading to modified spectra at large $p_T$. On the other hand, diffusion currents $V_X^\mu$ are vectors, and the lack of transverse flow in our simulations results in smaller effects on $p_T$ spectra. The difference in mean $p_T$ is less than 0.1\% for net baryons, and less than 1\% for net charges.

\section{Sensitivity of hadronic yields to conductivities}
\label{sec:sensitivity}
\vspace*{-2mm}

\subsection{Response surfaces}
\label{sec:surface}

We investigate the dependence of hadronic yields at midrapidity on the conserved charge conductivities. These quantities have been measured in the RHIC beam energy scan programs \cite{STAR:2008med,STAR:2017sal}. The sensitivities of the observables to the underlying transport coefficients are quantified through three approaches: (i) linear response matrix \cite{Sangaline:2015isa,Roch:2024xhh,Jahan:2024wpj,Gotz:2025wnv}, (ii) quadratic response tensor, and (iii) Gaussian process regression (GPR) \cite{Gong:2024lhq} models. The diagonal and off-diagonal conductivities are varied in the ranges of $\kappa_{BB}/T^2 \in [0,0.1]$, $\kappa_{BQ}/T^2 \in [0,0.005]$, $\kappa_{BS}/T^2 \in [-0.05,0]$, $\kappa_{QQ}/T^2 \in [0,0.05]$, $\kappa_{QS}/T^2 \in [0,0.02]$, $\kappa_{SS}/T^2 \in [0,0.25]$ to generate sample data sets of the multiplicities for $\pi^+$, $\pi^-$, $K^+$, $K^-$, $p$, $\bar{p}$, from which the response surfaces are constructed. The $\delta f$ corrections are included. We generate 400 data sets for $\sqrt{s_\mathrm{NN}}=19.6$, 62.4 GeV, 200 GeV from common sets of conductivities that satisfy positive semi-definiteness of $\mathcal{K}$. 

The linear response matrix $R$ is defined as
\begin{eqnarray}
\Delta \hat{N}_a=\sum_i R_{ai} \Delta K_i,
\end{eqnarray} 
where the hat symbol denotes prediction. Here $K_i = \kappa_i /T^2$ is the dimensionless conductivity and $\Delta K_i = K_i-\langle K_i \rangle $ is the deviation of the transport coefficient from its event average, where $i=\{ BB,BQ,BS,QQ,QS,SS\}$ is the index for the combination of charges.
$R$ is deduced from the data sets of the input $\Delta K_i$ and output $\Delta N_a$ of the hydrodynamic model using linear regression, where $\Delta N_a=dN_a/dy-\langle dN_a/dy \rangle $ is the deviation of hadronic yields from the average and $a=\{\pi^+, \pi^-, K^+, K^-, p, \bar{p} \}$ is the index for particle species. The prediction of the linear response matrix model $\Delta \hat{N}_a$ can be different from the true calculated $\Delta N_a$ in non-linear systems. The matrix elements can be expressed as $R_{ai}=\partial  (\widehat{dN_a/dy})/\partial K_i$. 

One can add the quadratic contributions by introducing the quadratic response tensor $Q$ defined as 
\begin{eqnarray}
\Delta \hat{N}_a=\sum_i R_{ai} \Delta K_i+\frac{1}{2} \sum_{i,j} Q_{aij} \Delta K_{i} \Delta K_{j},
\end{eqnarray} 
which is symmetric in $i$ and $j$, to take account of the non-linear dependencies on the conductivities. This leads to $Q_{aij}=\partial^2  (dN_a/dy)/\partial K_i\partial K_j$. 

Finally, we introduce the Gaussian process regression method following Ref.~\cite{Gong:2024lhq}. It is a non-parametric probabilistic model that can capture higher-order nonlinearities, which is widely used for machine learning. We use a radial basis function kernel assuming $dN/dy$ is a smooth function of the conductivities. 

\begin{table}[tb]
{   
\begin{tabular}{c|c|c|c|c|c|c|c}
\hline \hline
  & \multirow{2}{*}{methods} & \multicolumn{6}{c}{normalized RMSE (\%)} \\ \cline{3-8}
  & & $\pi^+$ & $\pi^-$ & $K^+$ & $K^-$ & $p$ & $\bar{p}$ \\ \hline
 & linear & 19 & 21 & 25 & 22 & 14 & 16 \\ \cline{2-8}
(i) & quadratic & 7.2 & 7.5 & 7.4 & 8.3 & 5.2 & 6.2 \\ \cline{2-8}
 & GPR & 0.29 & 0.41 & 0.39 & 0.33 & 0.10 & 0.15  \\ \hline
 & linear & 4.7 & 5.3 & 6.5 & 6.4 & 2.1 & 2.5 \\ \cline{2-8}
(ii) & quadratic & 2.0 & 2.5 & 3.0 & 2.4 & 1.1 & 1.1 \\ \cline{2-8}
 & GPR & 0.20 & 0.24 & 0.17 & 0.16 & 0.073 & 0.071  \\ \hline
 & linear & 3.6 & 2.0 & 2.2 & 3.5 & 5.4 & 5.7 \\ \cline{2-8}
(iii) & quadratic & 1.5 & 1.4 & 1.5 & 1.5 & 2.0 & 2.3 \\ \cline{2-8}
 & GPR & 0.18 & 0.20 & 0.15 & 0.14 & 0.088 & 0.085  \\ \hline\hline
\end{tabular}
\caption{The normalized RMSE of different response surfaces at (i) $\sqrt{s_\mathrm{NN}} = 19.6$ GeV, (ii) 62.4 GeV, and (iii) 200 GeV.} 
\label{table:rms}}
\end{table}

\begin{figure}[tb]
\includegraphics[width=3.3in]{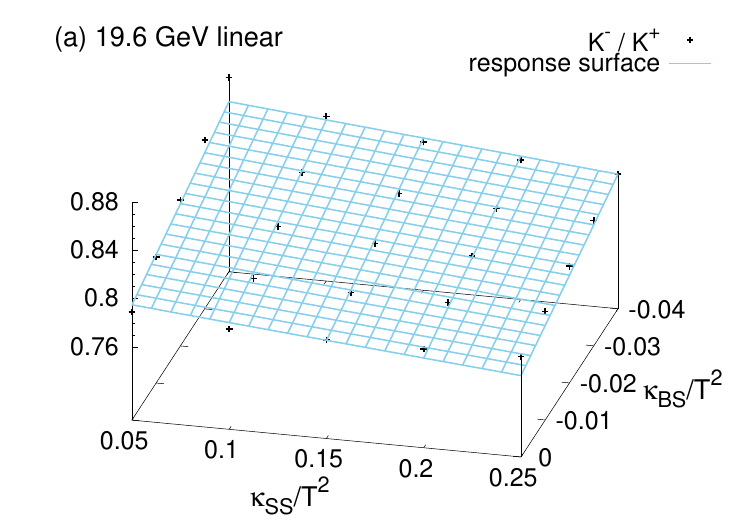}
\includegraphics[width=3.3in]{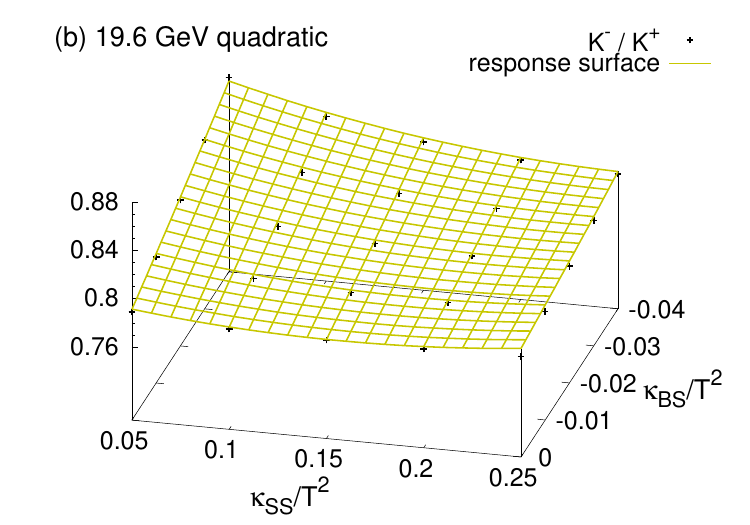}
\includegraphics[width=3.3in]{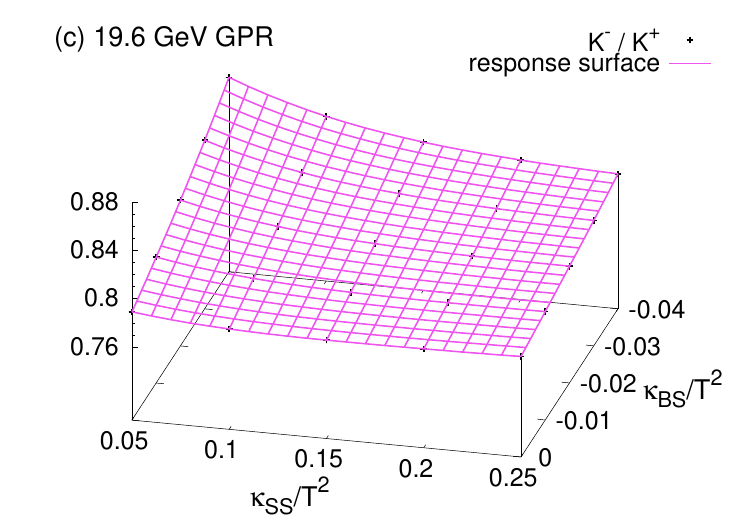}
\caption{The response surfaces of $K^-/K^+$ ratios based on the (a) linear response matrix, (b) quadratic response tensor, and (c) Gaussian process regression methods at $\sqrt{s_\mathrm{NN}} = 19.6$ GeV, compared with the ratio calculated directly from the hydrodynamic model.}
\label{fig:K_ratio}
\end{figure}

The normalized root mean square error (RMSE) from each method for the training data at different energies is summarized in Table~\ref{table:rms}. Here it is defined as the root mean square error divided by the standard deviation of the data:
\begin{equation}
\varepsilon_a = \frac{\sqrt{ \langle (dN_{a}/dy-\widehat{dN_{a}/dy})^2 \rangle}  }{  \sqrt{\langle(\Delta N_{a})^2\rangle}  }
\end{equation}
The precision improves from linear to quadratic approximations for all particles and collision energies. Non-linear effects tend to become more important at the lowest collision energies, possibly because the effects of diffusion currents are more pronounced. The GPR model provides a good description of the data in all cases. 

The antiparticle-particle yield ratios of kaons at $\sqrt{s_\mathrm{NN}} = 19.6$ GeV are plotted as functions of $\kappa_{SS}$ and $\kappa_{BS}$ in Fig.~\ref{fig:K_ratio} to illustrate the response surfaces from each method. The results of hydrodynamic simulations performed independently of the training data are plotted for comparison. The linear response matrix provides a reasonable approximation, though noticeable deviations appear in certain regions of the parameter space. Adding the quadratic response tensor significantly improves the agreement and reproduces the hydrodynamic results well. The Gaussian process regression approach further refines the description, accurately capturing the non-linear properties of the target variable.

\subsection{Experimental accessibility}
\label{sec:accessibility}

\begin{figure}[tb]
\includegraphics[width=3.3in]{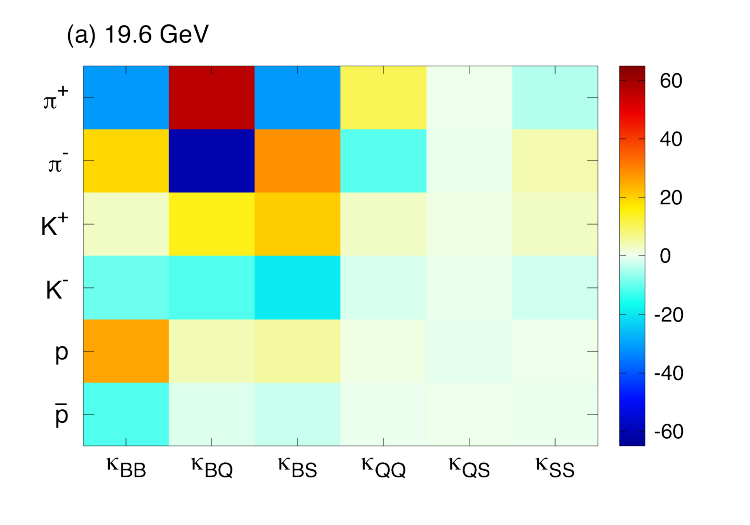}
\includegraphics[width=3.3in]{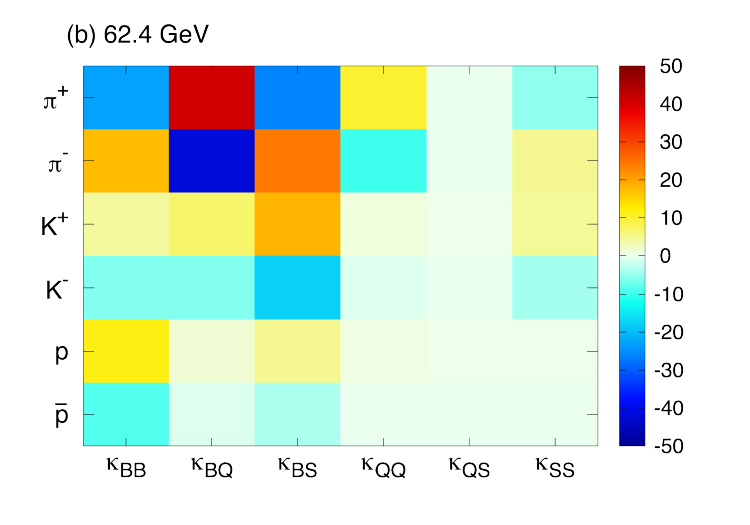}
\includegraphics[width=3.3in]{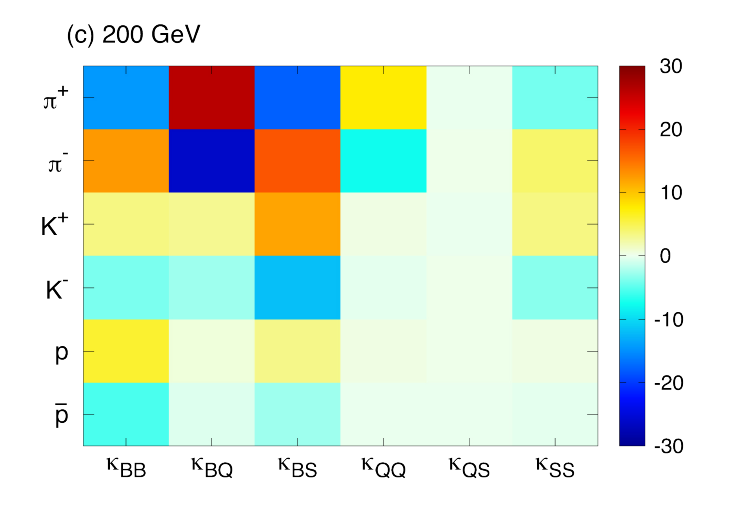}
\caption{The heatmap of $R_{ai} = \partial (\widehat{dN_a/dy})/\partial K_i$ at (a) $\sqrt{s_\mathrm{NN}} = 19.6$ GeV, (b) 62.4 GeV, and (c) 200 GeV, respectively.}
\label{fig:R_heatmap}
\end{figure}

Figure~\ref{fig:R_heatmap} is the heatmap of the linear response matrix $R$ at different collision energies. We find that, within the range of parameter exploration, $\pi^+$ and $\pi^-$ tend to receive the strongest influence from $\kappa_{BQ}$, followed by $\kappa_{BB}$ and $\kappa_{BS}$. This could be caused by the modification of the medium evolution and particlization hypersurface, which is consistent with the results of net charge distributions in Fig.~\ref{fig:dnqdy}. $K^+$ and $K^-$ are most sensitive to $\kappa_{BS}$, because the gradient of baryon chemical potential $\nabla_\mu (\mu_B/T)$ is larger than that of the strangeness chemical potential $\nabla_\mu (\mu_S/T)$ and the effect of the cross-coupling current is large. $\kappa_{SS}$ becomes relatively more important at higher energies. $p$ and $\bar{p}$ are primarily affected by $\kappa_{BB}$, seconded by $\kappa_{BS}$. $\kappa_{QS}$ has relatively small effects on all the hadronic yields at midrapidity as it characterizes the cross-coupling between minor diffusion currents. The effect of $\kappa_{QQ}$ is also small, except for that on pions.

One may consider a reverse process \cite{Sangaline:2015isa} to deduce the transport coefficients from the multiplicities of individual particles, characterized with the inverse response matrix $S$ defined as
\begin{eqnarray}
\Delta \hat{K}_i=\sum_a S_{ia} \Delta N_a,
\end{eqnarray} 
where $S_{ia}=\partial \hat{K}_i/\partial (dN_a/dy)$. In general, $S \neq R^{-1}$ because they are determined through the least squares method applied to different variables.

This simple approach, however, might encounter difficulties; although it is possible to formally define $S$ because $\Delta \mathbf{K}$ and $\Delta \mathbf{N}$ are both rank 6 vectors, the effects of diffusion on particles and antiparticles are typically similar in magnitude and opposite in sign, effectively reducing the rank of $\Delta \mathbf{N}$ to 3. Also, we have found that introducing additional particles such as $\Lambda$, $\Xi$, and $\Omega$ does not help solve the degeneracy issue, possibly because the particle numbers are controlled by the 3 chemical potentials $\mu_B$, $\mu_Q$, and $\mu_S$ on the particlization hypersurface. This makes the inverse response matrix unstable against small errors in multiplicity measurements.

The inversion problem could be mitigated by combining the results from different collision energies because of systematic differences in geometry. We define $S_\mathrm{global}$ as
\begin{eqnarray}
\Delta \hat{K}_i =\sum_{A}S_{\mathrm{global};iA} \Delta N_A,\label{eq:S_global}
\end{eqnarray} 
where $A=\{a_1,\dots , a_n \}$ is the index of the $n$ sets of the 6 particles.
For instance, in our estimations, the data at $\sqrt{s_\mathrm{NN}}=19.6$, 62.4 and 200 GeV would be labeled as $n=1,2$ and 3, respectively. It should be noted that if the sample datasets do not share a common $\langle K_i \rangle$ among different energies, the offsets need to be adjusted.

\begin{table}[tb]
{
\begin{tabular}{c|c|c|c|c|c|c}
\hline \hline
  \multirow{2}{*}{datasets} & \multicolumn{6}{c}{normalized RMSE (\%)} \\ \cline{2-7}
  & $\kappa_{BB}$ & $\kappa_{BQ}$ & $\kappa_{BS}$ & $\kappa_{QQ}$ & $\kappa_{QS}$ & $\kappa_{SS}$ \\ \hline
 (i) & 5.3 & 84 & 31 & 56 & 100 & 34  \\ \hline
 (ii) & 9.9 & 86 & 66 & 40 & 99 & 65 \\ \hline
 (iii) & 14 & 89 & 78 & 34 & 99 & 63 \\ \hline
 combined & 2.6 & 71 & 16 & 29 & 98 & 14  \\ 
\hline\hline
\end{tabular}
\caption{The normalized RMSE of the inverse linear response matrix at (i) $\sqrt{s_\mathrm{NN}} = 19.6$ GeV, (ii) 62.4 GeV, (iii) 200 GeV, and the combined dataset of all three energies.} 
\label{table:rms2}}
\end{table}

The normalized RMSE of the inverse linear response matrix, defined as
\begin{equation}
\varepsilon_i = \frac{\sqrt{ \langle (\kappa_i-\hat{\kappa}_i)^2 \rangle}  }{  \sqrt{\langle(\Delta \kappa_i)^2\rangle}  },
\end{equation}
from the fitting of the combined dataset of all three energies is compared with that of individual datasets in Table~\ref{table:rms2}. 
One can see that the local fits have difficulty in constraining the conductivities owing to the aforementioned degeneracy issue, except for $\kappa_{BB}$.
The accuracy increases when the global fitting is performed, particularly for $\kappa_{BS}$ and $\kappa_{SS}$, possibly because the collision energy dependence of strangeness diffusion is different from that of baryon and charge diffusion.
The results indicate that the inverse linear response matrix can be used to extract $\kappa_{BB}$ and can pose good constraints on $\kappa_{BS}$, $\kappa_{SS}$, and $\kappa_{QQ}$. On the other hand, it might be difficult to reliably constrain $\kappa_{BQ}$. $\kappa_{QS}$ is likely to be inaccessible from midrapidity yields since none of the $dN/dy$ observables depend on $\kappa_{QS}$ in our parameter settings.

Finally, we estimate the conductivities by solving a least-squares inverse problem using the response surface constructed with the GPR method. The normalized RMSE is shown in Table~\ref{table:rms3}. One can see that it is still difficult to constrain the transport coefficients from individual datasets. On the other hand, when multiplicities across different energies are used, the GPR method can determine the multicharge conductivities with improved accuracy, with the exception of $\kappa_{QS}$ to which the observables are insensitive. With this method, taking into account higher-order nonlinearities, $\kappa_{BQ}$ is experimentally accessible.

\begin{table}[tb]
{
\begin{tabular}{c|c|c|c|c|c|c}
\hline \hline
  \multirow{2}{*}{datasets} & \multicolumn{6}{c}{normalized RMSE (\%)} \\ \cline{2-7}
  & $\kappa_{BB}$ & $\kappa_{BQ}$ & $\kappa_{BS}$ & $\kappa_{QQ}$ & $\kappa_{QS}$ & $\kappa_{SS}$ \\ \hline
 (i) & 4.0 & 129 & 32 & 92 & 125 & 65  \\ \hline
 (ii) & 9.6 & 160 & 62 & 75 & 138 & 61 \\ \hline
 (iii) & 19 & 159 & 104 & 60 & 121 & 84 \\ \hline
 combined & 0.48 & 23 & 3.3 & 9.5 & 133 & 3.2 \\ 
\hline\hline
\end{tabular}
\caption{The normalized RMSE of the conductivities based on the GPR method at (i) $\sqrt{s_\mathrm{NN}} = 19.6$ GeV, (ii) 62.4 GeV, (iii) 200 GeV, and the combined dataset of all three energies.} 
\label{table:rms3}}
\end{table}

It should be noted that the purpose of the current study is to develop efficient methods to relate the conductivities and experimental observables, and a quantitative discussion, which requires full (3+1)-dimensional evolution and a hadronic afterburner, is left for future work. Our results are also subject to the choice of initial conditions. Nevertheless, the key conclusions regarding the qualitative sensitivities of $dN/dy$ to the conductivities should remain robust.

\section{Summary and outlook}
\label{sec:summary}
\vspace*{-2mm}

We have investigated baryon, electric charge, and strangeness diffusion effects in QCD matter based on a hydrodynamic model. We employed causal relativistic diffusion equations with cross-coupling terms. Off-equilibrium corrections to the phase space distribution $\delta f$ for multiple diffusion currents are implemented to ensure charge conservation at particlization. 

Hydrodynamic evolution in the longitudinal direction has been studied using numerical simulations of the most central Au+Au collisions at $\sqrt{s_\mathrm{NN}} = 19.6$ GeV, 62.4 GeV, and 200 GeV. Baryon diffusion is found to carry baryon number towards midrapidity in configuration space, and the effect becomes stronger at lower collision energies. The cross-coupling currents have, on the other hand, mostly negligible effects. Charge diffusion effects are small because $\mu_Q$ is much smaller than $\mu_B$ in magnitude. The off-diagonal terms also have marginal effects for the electric charge. The strangeness distribution develops a wave-like structure in the presence of strangeness diffusion even when the initial condition is neutral in strangeness because of the finite $\mu_S$ induced by $\mu_B$. The amplitude of this structure is affected by the cross-coupling currents. 

The flow is converted into particles based on the Cooper-Frye method supplemented with the aforementioned distortion of the momentum distribution, $\delta f$. Baryon diffusion leads to increased baryon density around midrapidity, with the effect becoming larger at lower energies. $\delta f$ corrections are found to visibly modify $dN_B/dy$. $dN_Q/dy$ is also affected by the diffusion current but to a lesser extent, similarly to its spacetime counterpart. The effects of the cross-coupling currents and the off-equilibrium distributions are limited.
$dN_S/dy$ is non-vanishing even in the ideal case because of the difference between the thermal smearing of baryons and mesons in momentum space. Strangeness diffusion induces a characteristic wave-like structure on top of that, and is sensitive to the off-diagonal currents. The $\delta f$ correction also visibly modifies the distribution and plays an essential role in preserving global strangeness neutrality.

We have investigated the sensitivity of $dN/dy$ at midrapidity to the conductivities to develop a method to extract the transport coefficients from the quantities measured in the RHIC BES programs \cite{STAR:2008med,STAR:2017sal}. The response surfaces are constructed based on the linear response matrix, the quadratic response tensor, and Gaussian process regression methods. While the linear response matrix provides a reasonable description of the particle yields, it is outperformed by the other two methods, especially at lower energies where nonlinearities become more relevant. The GPR accurately describes multiplicities for all particles at all energies.

The heatmap of the response matrix implies that pions are sensitive to $\kappa_{BB}$, $\kappa_{BQ}$, and $\kappa_{BS}$, kaons to $\kappa_{BS}$, $\kappa_{SS}$, and $\kappa_{BQ}$, and protons to $\kappa_{BB}$ and $\kappa_{BS}$. We have combined the results from different collision energies to overcome the degeneracy issue, which means there are 6 conductivities for 3 conserved charges. It has been found that the global inverse response matrix can be used as a simple way to extract $\kappa_{BB}$ and constrain $\kappa_{BS}$, $\kappa_{SS}$, and $\kappa_{QQ}$ with reasonable accuracy. We have also shown that the nonlinear response surfaces obtained from the GPR method can be a powerful tool to determine the QCD conductivities, including $\kappa_{BQ}$, with much improved accuracy. On the other hand, weak sensitivity makes it difficult to constrain $\kappa_{QS}$ from the multiplicities alone. 

Our future plans include a full (3+1)-dimensional hydrodynamic analysis with a hadronic afterburner, which is necessary for quantitative analyses. The temperature dependence of the charge conductivities beyond $T^2$ might be constrained by introducing observables other than net charges at midrapidity. The mean rapidity shift of conserved charges will be sensitive to the conductivities. Also, flow harmonics $v_n$ might provide additional information \cite{Denicol:2018wdp}. It would also be interesting to consider initial conditions with different baryon and charge stopping, or with fluctuating strangeness density. Our numerical results for $\delta f$ will be made publicly available \cite{deltaf}.

\begin{acknowledgments}
This work is supported by JSPS KAKENHI Grant Number JP24K07030 (A.M.), by the U.S. Department of Energy, Office of Science, Office of Nuclear Physics, under DOE Contract No.~DE-SC0012704 and within the framework of the Saturated Glue (SURGE) Topical Theory Collaboration (B.P.S.) and Award No.~DE-SC0021969 (C.S. \& G.P.). C.S. acknowledges a DOE Office of Science Early Career Award. 
\end{acknowledgments}

\bibliography{diffusion}

\end{document}